\documentclass[a4paper,twocolumn]{article}
\usepackage[english]{babel}

\usepackage{graphicx}

\usepackage{caption}
\usepackage{subcaption}
\newcommand{\BN}{\textit{h}-BN}

\begin{document}

\title{Phonons and electron-phonon coupling in graphene-h-BN heterostructures}
\author{Guus J. Slotman\footnote{Institute for Molecules and Materials, Radboud University Nijmegen}~\footnote{gslotman@science.ru.nl}, Gilles A. de Wijs\footnotemark[1],\\ Annalisa Fasolino\footnotemark[1]~ and Mikhail I. Katsnelson\footnotemark[1]}

\maketitle

\begin{abstract}
First principle calculations of the phonons of graphene-\BN~ heterostructures are presented and compared to those of the constituents. We show that AA and AB' stacking are not only energetically less favoured than AB but also dynamically unstable.  We have identified low energy flat phonon branches of \BN~character with out of plane displacement and evaluated their coupling to electrons in graphene. 
\end{abstract}



\section{Introduction}
After the discovery of graphene, several other two-dimensional crystals were created and their combination in what have been called van der Waals heterostructures is currently in the focus both for fundamental science and potential applications\cite{geim2013van}.   Hexagonal boron nitride (\BN) is particularly important as the best known insulating substrate for graphene to reach the highest electron mobility\cite{dean2010boron,dean2011multicomponent,tang2013precisely,gannett2011boron,
decker2011local,mayorov2011micrometer,ponomarenko2011tunable, castro2010limits}. 
The small lattice mismatch of 1.8\% between \BN~ and graphene and the possibility to change the relative orientation provide a tunable modulation of the lattice. This provides a playground to study electronic structure and transport in incommensurate potentials, one of the fundamental issues in quantum mechanics \cite{yankowitz2012emergence, ponomarenko2013cloning, dean2013hofstadter, hunt2013massive,woods2014commensurate, xue2011scanning}.
The first graphene-\BN~field-effect transistor\cite{britnell2012field} based on tunneling of electrons from graphene through \BN~ was recently created,  paving the way for other graphene-\BN~devices\cite{bresnehan2012integration,yankowitz2014graphene}.
The interface between graphene and \BN~is therefore interesting both for structural and electronic properties. The phonon spectrum of the combined structure is important for characterization and the coupling to electrons can affect tunneling characteristics. 

In this article we calculate from first principles the phonon spectra of the graphene-\BN~heterostructure in comparison to the constituents. We found several flat phonon branches of \BN~character with out of plane displacement that provide large peaks in the phonon density of states and  evaluate their coupling to electrons in graphene. Many theoretical  papers studying graphene-\BN~systems have been published\cite{giovannetti2007substrate, zhong2012electronic, bokdam2011electrostatic, sachs2011adhesion, titov2014metal, bokdam2014band} and although \textit{ab initio} calculations of the phonon spectra exist for both graphite\cite{mounet2005first} and BN\cite{kern1999ab}, to our knowledge phonons of combined structures have not yet been reported in the literature. 

In section~\ref{sec:computationalmethods} we describe the computational approach and in section~\ref{sec:structure} the geometry of the interface. In ~\ref{sec:phonons} the calculated  phonon spectra are presented and in~\ref{sec:electronphonon} we present our results on  electron phonon coupling. A summary and conclusions are given in~\ref{sec:conclusion}.

\section{Computational Methods}\label{sec:computationalmethods}
First principles calculations of the electronic structure and the phonon dispersions were performed using the Vienna \textit{ab initio} simulation package (\textsc{vasp})\cite{kresse1996vasp1,kresse1996vasp2}.
As standard DFT (eg. LDA, GGA) fails in accurately describing the inter-layer van der Waals interactions, multiple schemes have been proposed to deal with dispersive forces. We use the van der Waals density functional proposed by Dion et al.\cite{dion2004van} (vdW-DF) as implemented in \textsc{vasp} by Klime\v{s} et al.\cite{klimes2010chemical,klimes2011van} using the algorithm of Rom\'an-Pe\'rez and Soler\cite{roman2009efficient}. 
One can combine this vdW-DF scheme with various functionals. The original paper\cite{dion2004van} uses the revPBE functional, but Hazrati et al. \cite{hazrati2013phd} have shown that the optB88 describes better the inter-layer distance of graphite. The optB88 functional was also used by others to study the \BN-metal interface \cite{bokdam2014schottky}. Therefore we use the optB88 functional for all phonon calculations.
The standard PAW data sets of \textsc{vasp} were used\cite{blochl1994projector, kresse1999ultrasoft}. The cutoff energy for the planewave basis set was $550$ eV. A second order Methfessel-Paxton scheme \cite{methfessel1989high} with a smearing of $0.2$ eV was used to determine the occupancy of the electronic levels. While doing band-structure calculations with the optB88-vdW functional we noticed however that there were large fluctuations of the conduction bands as a function of the box-size (eg. the amount of vacuum) and therefore we used the optB86-vdW functional for the band structure calculations of graphene and \BN~systems to evaluate the electron phonon coupling.

For graphene-\BN~heterostructures, we distinguish between bulk and bilayer systems: the bulk system consists of one layer of \BN~and one of graphene, periodically repeated in all directions, while the bilayer has a vacuum of $\approx 25$ \AA~ separating the periodic images in the z-direction. For comparison we also calculate the phonon spectra of single layer and bilayer graphene and of single layer and bilayer \BN. A $\Gamma$-centered $24 \times 24 \times 6$ $k$-point mesh was used for bulk systems, and a $24 \times 24 \times 1$ for bilayer systems.  For the phonon calculations we used the same $k$-point densities in the suitable supercells described later. 
The dynamical matrix for the phonon dispersion was calculated by the finite differences method\cite{kresse1995ab,parlinski1997first,phonopy}, as implemented by\cite{phonopy}, by displacing the atoms by $\pm0.015$~\AA~in each symmetrically non-equivalent direction. To improve accuracy, the initial equilibrium atomic positions were relaxed until the forces were less than $0.0001$~eV$/$\AA.

\section{Structure}\label{sec:structure}

\begin{figure}[ht!]
\begin{center}
\includegraphics[angle=0,width=0.49\textwidth]{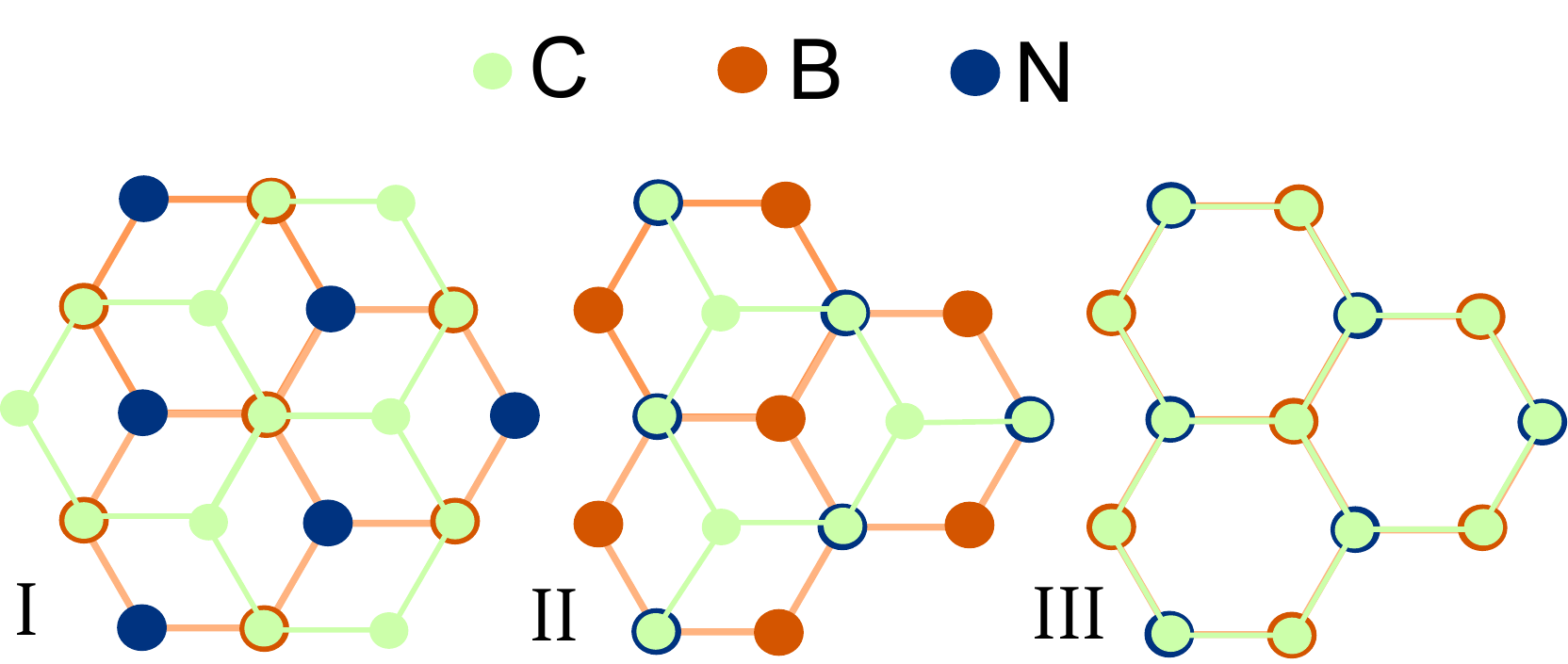}
\end{center}
\caption{The three stackings we consider for C-\BN~systems. (I) The AB-stacking, with a C atom on top of a B atom, is energetically the most favourable. (II) AB$'$ with a C atom on top of a N atom. (III) AA-stacking.  }
\label{fig:stackings}
\end{figure}
\begin{figure}[th!]
\begin{center}
\includegraphics[angle=0,width=0.39\textwidth]{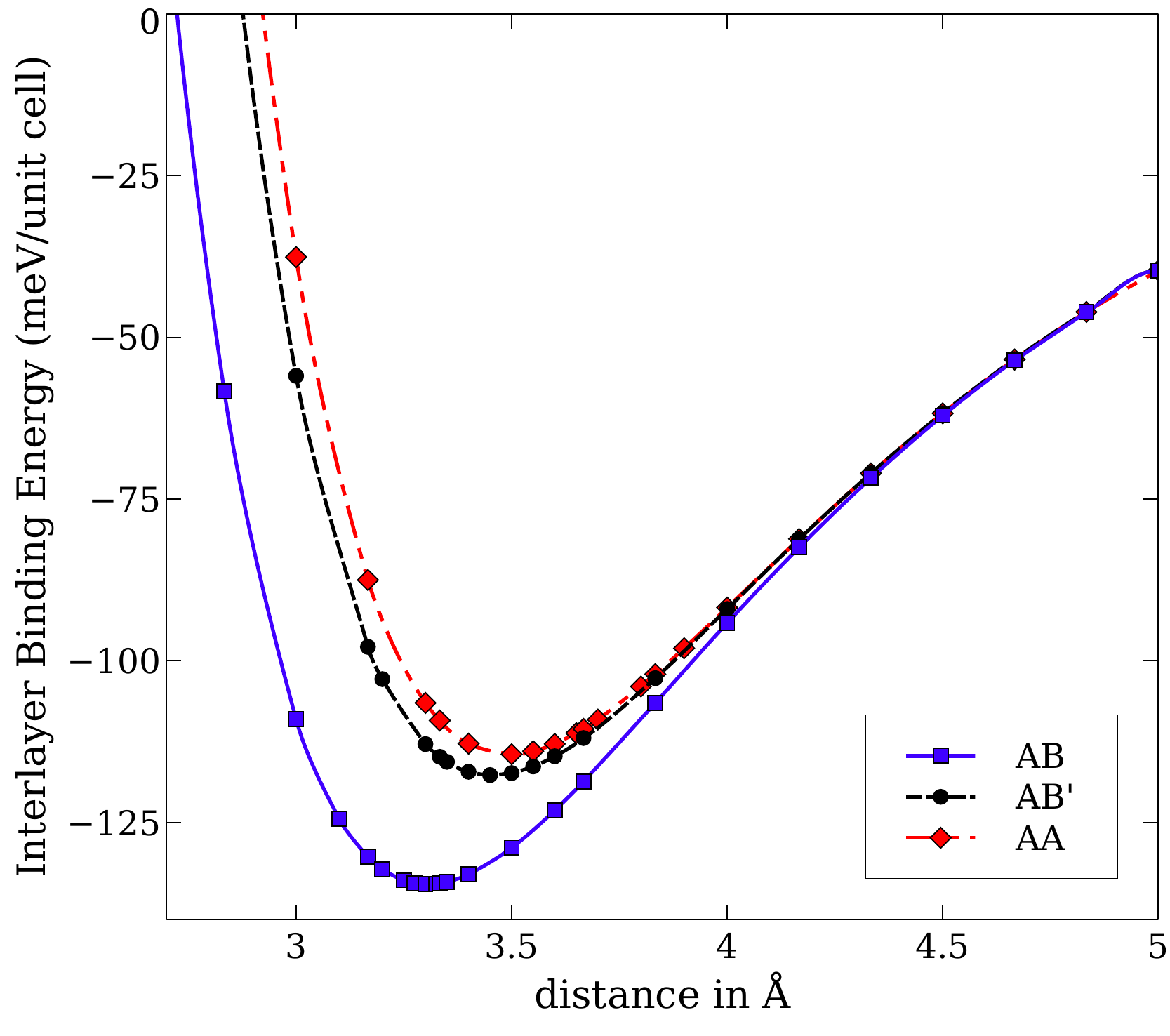}
\end{center}
\caption{Interlayer binding energy per unit cell of a graphene-\BN~bilayer as a function of the distance between the two layers for in-plane lattice constant  $a=2.49$ \r{A}. The zero corresponds to the energy calculated at a separation of 15 \AA. }
\label{fig:lattice_para}
\end{figure}

The lattice parameter of \BN~is  $1.8\%$ larger than that of graphene, leading to moir\'e patterns \cite{hermann2012}, which are also observed experimentally \cite{decker2011local, xue2011scanning, yankowitz2012emergence, hunt2013massive}. Therefore the description of graphene-\BN~requires large unit cells which are beyond the reach of DFT methods.

Our strategy has been to use the normal bulk unit cell (two atoms per layer) and we determined, by minimizing the energy, the lattice parameter $a=2.49$~\AA~and interlayer distance $d=3.30$~\AA~for the combined structure. In the case of bulk materials $d$ is half of the out of plane lattice vector $c$. These values can be compared with the ones we find for graphite and bulk \BN, $a=2.47$~\AA~ $d=3.36$~\AA~ and $a=2.505$~\AA~ $d=3.30$~\AA~, respectively. Further accuracy was obtained by minimising the residual forces until they were less than $0.0001$~eV$/$\AA.

The various stackings studied are shown in figure~\ref{fig:stackings}. As shown in various calculations\cite{giovannetti2007substrate,sachs2011adhesion}, the energetically most favourable configuration for combined graphene-\BN~systems is the AB stacking where one carbon atom sits on top of a boron atom and one sits in the middle of a BN hexagon. The other stackings are the AB$'$ where one carbon atom sits on top of one nitrogen atom and the AA stacking. For bulk \BN,  the AA stacking corresponds to two possible situations, where the one with a boron atom on top of a nitrogen atom is the most stable. For graphite the AB is the most favourable stacking.

In figure~\ref{fig:lattice_para} we show the interlayer binding energy for a graphene-\BN~bilayer as a function of the distance between the two layers for the different stackings. The AB stacking (I in fig.~\ref{fig:lattice_para}) is indeed the most stable configuration at a interlayer distance $d=3.30$~\AA~($-135$~meV). The results can be compared to more accurate ACFDT-RPA calculations done by Sachs et al.\cite{sachs2011adhesion}, who found an interlayer distance of $3.35$~\AA~with a binding energy of $-83$ meV, meaning that our binding energy is $\approx1.6$ times larger. The overestimation of the binding energy by the vdW functional is also observed in other studies where the interaction between small molecules and graphene is studied\cite{hazrati2013phd, hamada2012adsorption, silvestrelli2014including}. For the AB$'$  stacking we find $d=3.45$~\AA~ ($-118$~meV) and for the AA stacking  $d=3.49$~\AA~ ($-114$~meV).

\section{Phonons}\label{sec:phonons}

\begin{figure}[ht!]
        \centering
        \begin{subfigure}[b]{0.49\textwidth}
                \includegraphics[width=\textwidth]{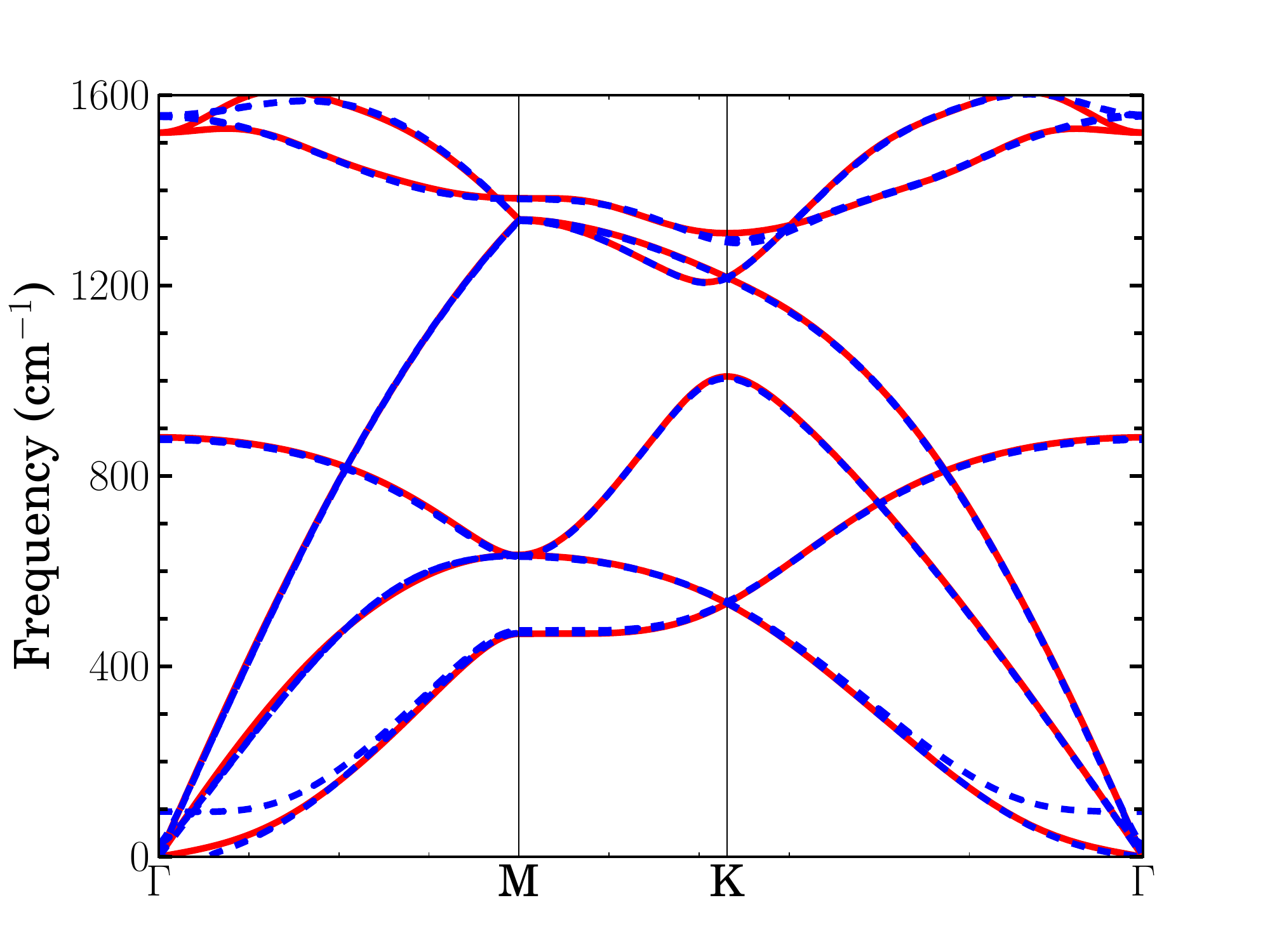}
                \caption{Graphene single-layer (red line) and bilayer (blue dashed)}
                \label{fig:c_b_phon_c}
        \end{subfigure}\\
        \begin{subfigure}[b]{0.49\textwidth}
                \includegraphics[width=\textwidth]{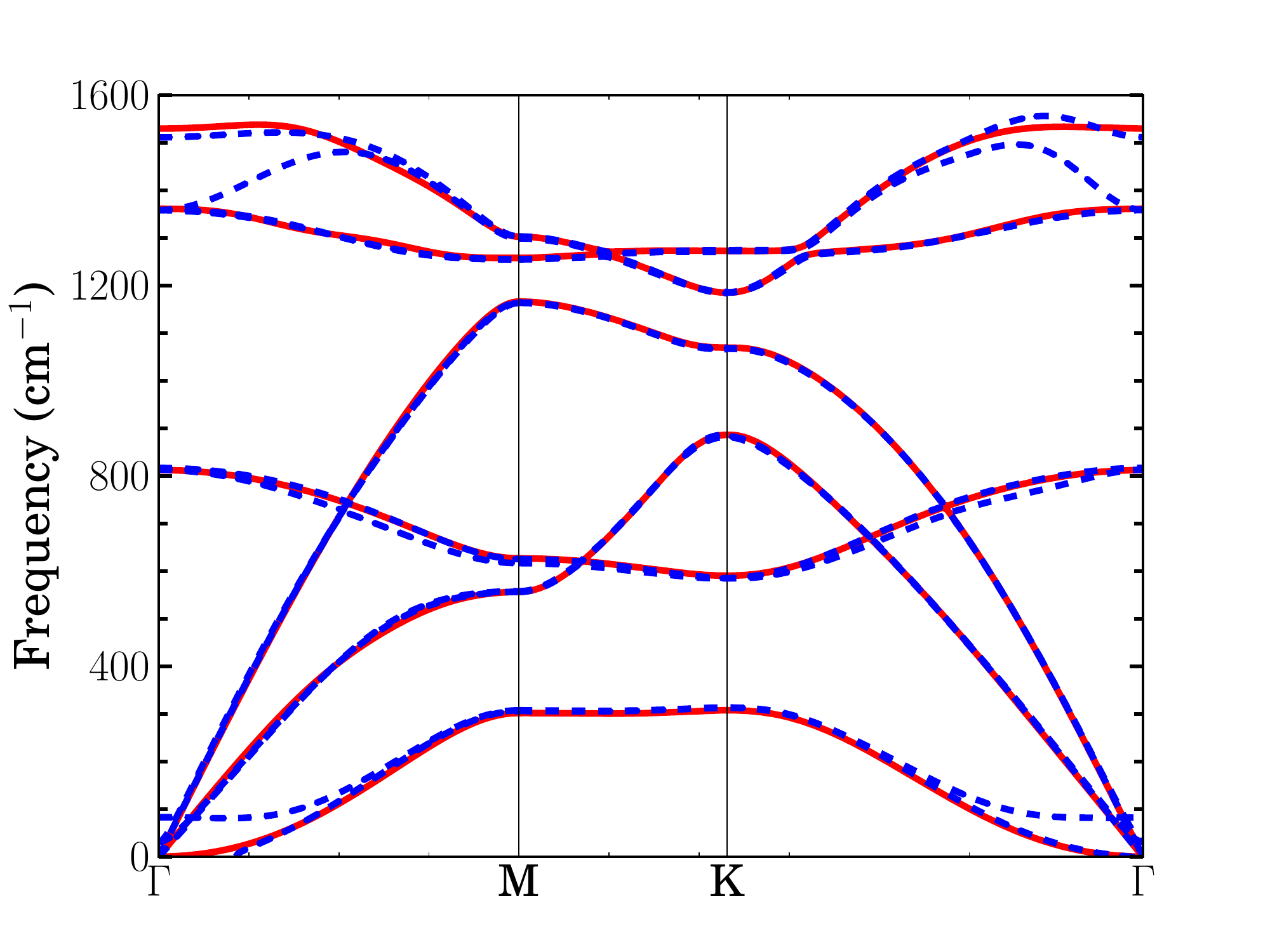}
                \caption{\BN~single-layer (red line) and bilayer (blue dashed)}
                \label{fig:c_b_phon_bn}
        \end{subfigure}%
        \caption{The phonon spectra of single-layer and bilayer (a) graphene and (b) \BN . For \BN~a Gaussian smearing and a non-analytical correction were used to account for the LO-TO splitting at $q \to \Gamma$ (see e.g. Ref. \cite{baroni2001}). For both bi-layer systems we used the energetically most favourable stackings (AB for graphene and AA$'$ for \BN).}\label{fig:c_b_phon}
\end{figure}

We first calculate the phonon spectra of graphene (figure~\ref{fig:c_b_phon_c}) and \textit{h}-BN (figure~\ref{fig:c_b_phon_bn}) for both single layer and bilayer systems. 
For graphene, $a=2.47$~\AA~ and the interlayer distance for the bilayer system is $d =3.36$~\AA; for \BN~$a=2.505$~\AA~ and $d=3.35$~\AA.  
For the phonon calculations we used a $4 \times 4 \times 1$ supercell for the $\Gamma$-M line in the Brillouin-zone and a $8a \times \sqrt{3}a \times 1$ cell for  M-K-$\Gamma$. 
For the single layer calculations we used a $8 \times 8 \times 1$ supercell because smaller cells gave rise to small imaginary frequencies ($\omega^2 < 0$) close to $\Gamma$. 

The main differences between the phonon dispersions  of graphene and \BN~are in the lowest lying modes. For carbon the lowest mode at K is fourfold degenerate while for \BN~this degeneracy is lifted due to the two different atoms in the unit cell. As a result, in \BN~the lowest two branches, which are both polarized out of plane, are almost flat over the whole M-K range. The eigenvector of the lowest branch involves mostly the nitrogen atom and that of the second branch mostly the boron atom. 
The frequencies of these modes are given in table~\ref{table:mk_values}. 

\begin{figure}[h!]
\begin{center}
\includegraphics[angle=0,width=0.49\textwidth]{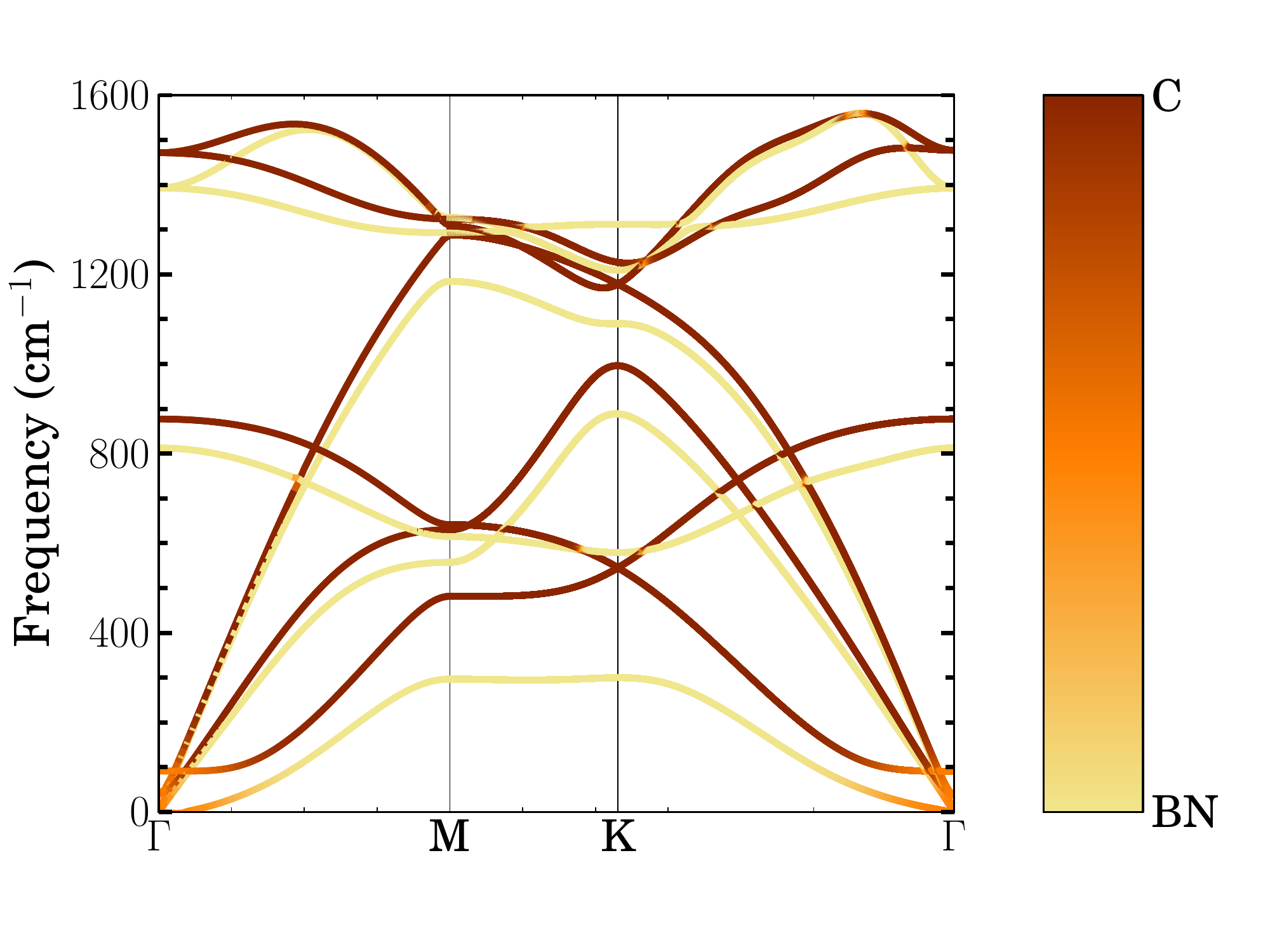}
\end{center}
\caption{Phonon dispersion of a graphene-\BN~bilayer, calculated with the optimized lattice parameter $a=2.49$~\r{A} and interplanar distance  $d=3.30$~\r{A}).  The colours show which layer contributes the most to a particular mode. }
\label{fig:bnc_phonon_vac}
\end{figure}

In figure~\ref{fig:bnc_phonon_vac} we show the phonon dispersion of a bilayer graphene-\BN. 
The spectrum is almost the superposition of the spectra of the two constituents. 
In the figure the colours represent the contributions of each layer to the phonon mode. Each mode is rather localized on either layer with the exception of the acoustical modes around $\Gamma$.
This coupling makes that the two lowest acoustical branches at $\Gamma$ anticross and reach M at different frequencies. These modes remain almost flat between M and K. We will discuss the coupling to electrons of these flat branches in section~\ref{sec:electronphonon}. 

It is worth noting that there is a frequency ratio of roughly 1 to 2 of the two lowest lying \BN-branches with z-polarization which can result in the non-perturbative anharmonic effect known as Fermi resonance. This can lead to new phenomena like splitting of the upper branch and low frequency dissipation associated with energy transfer between the two modes.\cite{fermi1931ramaneffekt, fermi1966molecules, lisitsa1984, katsnelson2004nonperturbative}

\begin{figure}[th!]
\begin{center}
                \includegraphics[angle=0,width=0.49\textwidth]{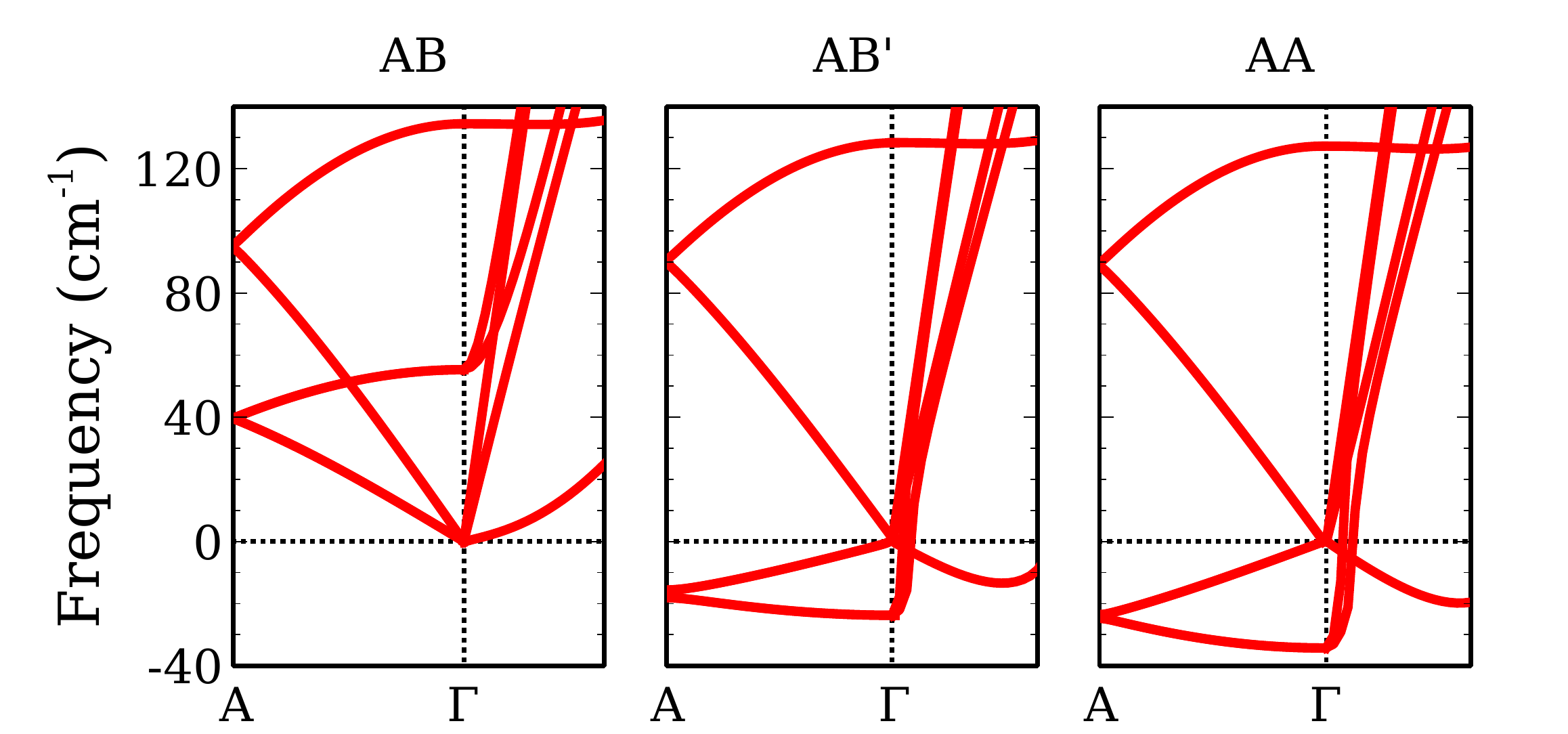}
\end{center}
\caption{Low frequency phonon dispersion relations along $A-\Gamma$ and from $\Gamma$-$^1/_3$M for the three different stackings for bulk graphene-\BN. The AB$'$ and AA stackings have negative (imaginary) frequencies.For higher frequencies the different stackings yields the same result.
}
\label{fig:stack_phonon}
\end{figure}

In figure \ref{fig:stack_phonon} we show the phonon dispersion of bulk graphene-\BN~ along the line A-$\Gamma$ and from $\Gamma$
to $^1/_3$M for the three different stackings.
For the  A-$\Gamma$ calculations the unit cell was doubled in the z-direction. We find that the transverse in-plane modes have
imaginary frequencies ($\omega^2<0$) for the less stable AB$'$ and AA configurations over the whole A-$\Gamma$ line, meaning that the layers tend to slide onto each other. This means that the AB$'$ and AA configurations are not only energetically less favourable but also dynamically unstable.

Note that, for the unstable stackings, there is also a smaller imaginary part of the z-polarized flexural mode in the line $\Gamma$-M. This is a numerical error related to the supercell size analogous to the one discussed for a single layer due to the peculiar quadratic dispersion of the flexural modes. These imaginary frequencies are smaller and become real well before reaching the M point. 

\begin{table*}[t]
  \begin{tabular}{l | c c c c c c } 
 &  &  & M & & &\\
 \hline
 \BN &307.1& - & 625.6 & - &  1255.3 & - \\
 C & - & 472.7 & - &632.0 & - & 1382.3 \\
 C-\BN &296.8 & 481.4 & 614.9 & 641.4 & 1292.8& 1323.5 \\
 \hline
&  &  & \textbf{K} & & &\\
\hline
\BN & 313.4 (g=0.24) & -  &585.1 (g=0.16) & - &1272.6 & - \\
C & -& 534.0 (g=0.048)&-  &- & -& 1294.4\\
C-\BN& 299.8 & 545.1 & 579.1  & - &1311.8 & 1226.9 \\
  \end{tabular}\caption{Frequencies for the flat bands at M and K in cm$^{-1}$. Except the last two columns, which are in plane modes, all values listed are for out of plane modes. $1$ cm$^{-1}=8.1^{-1}$~meV. For selected modes we give the e$_{ph}$ coupling constant g\label{table:mk_values} between brackets.}%
\end{table*}

\section{Electron-phonon coupling}\label{sec:electronphonon}
 
The peculiar phonon spectrum of graphene-\BN~discussed in the previous section, with low energy flat branches with out of plane displacement of the nitrogen or of the boron atom lead to  enhanced van Hove singularities. In two dimensional crystals, in general, the van Hove singularities in the phonon density of states are either finite discontinuities (at the edges of the phonon dispersion) or logarithmic divergencies at saddle points~\cite{vanhove1953occurrence}.
The lowest branch between M and K  that connects the top of the flexural acoustic modes along $\Gamma$-M to the K point has a dispersion of only  0.3 meV, making the spectrum effectively one dimensional. In this situation, one can expect a much stronger van Hove singularity $\propto \left(\omega_c-\omega\right)^{-1/2}$ ($\omega < \omega_c$) typical of the one dimensional case. This strong van Hove singularity should be visible in IV characteristics of tunnelling currents (see e.g. Refs. \cite{stipe1998, smit2002, wehling2008}). Along the M-K direction there are also other branches with little dispersion at higher energies, for which the values are also  given in table~\ref{table:mk_values}. An analysis of the phonon eigenvectors at the K point shows that the branches with $E=299.8$~cm$^{-1}$ ($37.1$~meV) and $E=579.1$~cm$^{-1}$  ($71.8$~meV) at K correspond to out of plane displacement of the nitrogen and boron atoms  respectively whereas the highest flat branch correspond to in-plane motion. Obviously out of plane modes of \BN~should interact much more strongly with graphene than in-plane modes. 

Phonons at the K point connect the K and K' electron valleys in graphene\cite{katsnelson2012graphene} leading to a splitting $\Delta E = 2|H_{KK'}|$ where $H_{KK'}$ is the matrix element of the Hamiltonian in the lattice distorted according to the phonon eigenvectors at K and K'. For a single \BN~layer, single phonon processes are forbidden by the symmetry $z\rightarrow -z$ and one  would have to consider two phonons processes where the splitting should  be proportional to the square of the phonon displacements. For the bilayer BN instead, single phonon processes are allowed due to the modulation of the interlayer hopping parameters. 

To calculate the dimensionless electron-phonon coupling constant $g$, defined as $g=\sqrt{\frac{\hbar}{2M \omega}} \frac{1}{\hbar \omega} \frac{\Delta E}{ \Delta u}$ we make an estimate for $\frac{\Delta E}{ \Delta u}$ by use of a frozen phonon displacement.
Since the modes we consider correspond to the displacement of a single type of atom, the mass $M$ is taken as that of the relevant atom. Atoms are displaced according to the eigenvector $u$ of the considered phonon mode. The displacements break the symmetry of the lattice and the points $K$ and $K'$ are no longer equivalent. $E$ is then taken to be the difference in energy between the top of the valence band at $K$ and at $K'$. 
In this way, we have calculated the values of $g$ given in table~\ref{table:mk_values} for the low lying out of plane modes for both bilayer \BN~and graphene. We see that the electron phonon coupling is strongest for the mode with \BN~character. The mode dominated by the nitrogen atoms have the highest $g$. This fact might have been expected because the van der Waals interactions between carbon  and nitrogen have been found to be stronger than those between carbon and boron\cite{sachs2011adhesion}.

\section{Conclusion}\label{sec:conclusion}
We have shown that the phonon dispersions of graphene-\BN~heterostructures  are very close to the superposition  of the ones of the constituent layers, with mixing only for the acoustical modes near $\Gamma$. The phonon dispersions, reveal a dynamical instability of the less favourable stackings. Flat branches with \BN~character are shown to provide a sizeable electron phonon coupling to the valence band of graphene.

\section*{Acknowledgements}
We thank K. Novoselov, E. Hazrati and M. Marsman for discussions. This work is part of the research program of the Foundation for Fundamental Research on Matter (FOM), which is part of the Netherlands Organisation for Scientific Research (NWO). The research leading to these results has received funding from the European Union Seventh Framework Programme under grant agreement No. 604391 Graphene Flagship and was supported by the  ERC Advanced Grant No. 338957 FEMTO/NANO.


\end{document}